**A Self-Healing Framework for Reliable LLM-Based Autonomous Agents**

Cheonsu Jeong*

*AX Center, SAMSUNG SDS, Seoul, South Korea*
*paripal@korea.ac.kr*

Younggun Shin

*Dept. of Artificial Intelligence, Graduate School of Engineering, Yonsei University, Seoul, South Korea*
*ygworld50@yonsei.ac.kr*

Autonomous agents based on Large Language Models (LLMs) are increasingly being utilized in complex software systems. However, reliability remains a significant challenge due to unpredictable failures such as hallucinations, execution errors, and inconsistent reasoning. This paper proposes a reliability-aware self-healing framework for LLM-based software agents. The framework integrates failure detection, reliability assessment, and automated recovery mechanisms. First, we define a taxonomy of failure types and introduce a quantitative reliability assessment model. Next, we propose a failure detection method that identifies abnormal agent behavior based on execution patterns and output consistency. Finally, we design a self-healing mechanism that dynamically recovers from failures through adaptive replanning and corrective prompting strategies. The proposed framework was implemented in a multi-agent workflow environment and evaluated using real-world task scenarios. Experimental results demonstrate that our approach significantly increases task success rates, reduces failure propagation, and enhances overall system robustness compared to existing methods. In particular, this study distinguishes itself by establishing an integrated monitoring system that combines the agent's internal reasoning process with external execution results. These findings are expected to contribute to securing the stability of advanced autonomous systems and lowering the barriers to LLM adoption in production environments.

## 1. Introduction

Recent advances in Large Language Models (LLMs) have enabled the development of autonomous software agents capable of performing complex tasks such as reasoning, planning, and tool usage. These agents have been increasingly applied to real-world scenarios, including workflow automation and multi-agent collaboration systems [1]. In particular, approaches such as ReAct [2] and HuggingGPT [3] demonstrate how LLMs can coordinate reasoning and action to accomplish multi-step tasks.

Despite these advances, LLM-based agents still suffer from significant reliability issues. Common failure modes include hallucinated outputs, incorrect tool usage, and inconsistent reasoning processes [4]. These issues become more critical in multi-step workflows, where errors can propagate and amplify across task execution stages.

Existing research has attempted to improve reliability through techniques such as prompt engineering, self-consistency decoding, and iterative refinement [5, 6]. While these approaches improve output quality to some extent, they are primarily designed for static inference settings and do not effectively address runtime failures in dynamic environments.

* Corresponding author: csu.jeong@samsung.com

From a software engineering perspective, reliability is not only about correctness but also about robustness and recoverability. Traditional autonomic computing systems emphasize self-monitoring and self-healing capabilities to maintain system stability under uncertain conditions [7, 8]. Recently, the integration of LLMs into self-adaptive multi-agent systems has shown promise in enhancing system resilience [9, 10]. However, a comprehensive framework that unifies failure detection, quantitative reliability evaluation, and automated recovery specifically tailored for LLM-based agents remains underexplored.

To address these limitations, this paper proposes a reliability-aware self-healing framework for LLM-based autonomous software agents. The proposed framework integrates reliability evaluation, failure detection, and automated recovery into a unified system, drawing inspiration from both modern AI agent architectures and traditional autonomic computing principles.

The main contributions of this paper are as follows:

- A quantitative reliability evaluation model for LLM-based agents that operates dynamically during runtime.
- A hybrid failure detection mechanism based on execution patterns and output consistency.
- A self-healing framework that enables automatic recovery through adaptive strategies such as re-planning and prompt correction.

The effectiveness of the proposed approach is validated through experiments in multi-agent workflow scenarios, demonstrating significant improvements in task success rates and system robustness.

## 2. Related Work

This chapter reviews existing studies related to LLM-based autonomous agents, reliability enhancement, and self-healing mechanisms. As illustrated in Fig. 1, prior research can be broadly categorized into three directions: LLM-based autonomous agent systems, reliability and failure handling techniques, and self-healing or adaptive recovery approaches.

Although these studies have significantly advanced the capabilities of LLM-based systems, they are typically developed in isolation. In particular, existing approaches tend to focus on specific aspects such as task execution, output quality improvement, or partial recovery strategies, without providing a unified mechanism that covers the entire lifecycle of agent execution.

As highlighted in Fig. 1, the key limitation of existing research is that reliability evaluation, failure detection, and recovery mechanisms are treated as separate components. This fragmented design restricts their effectiveness in real-world environments, where failures occur dynamically and require coordinated responses.

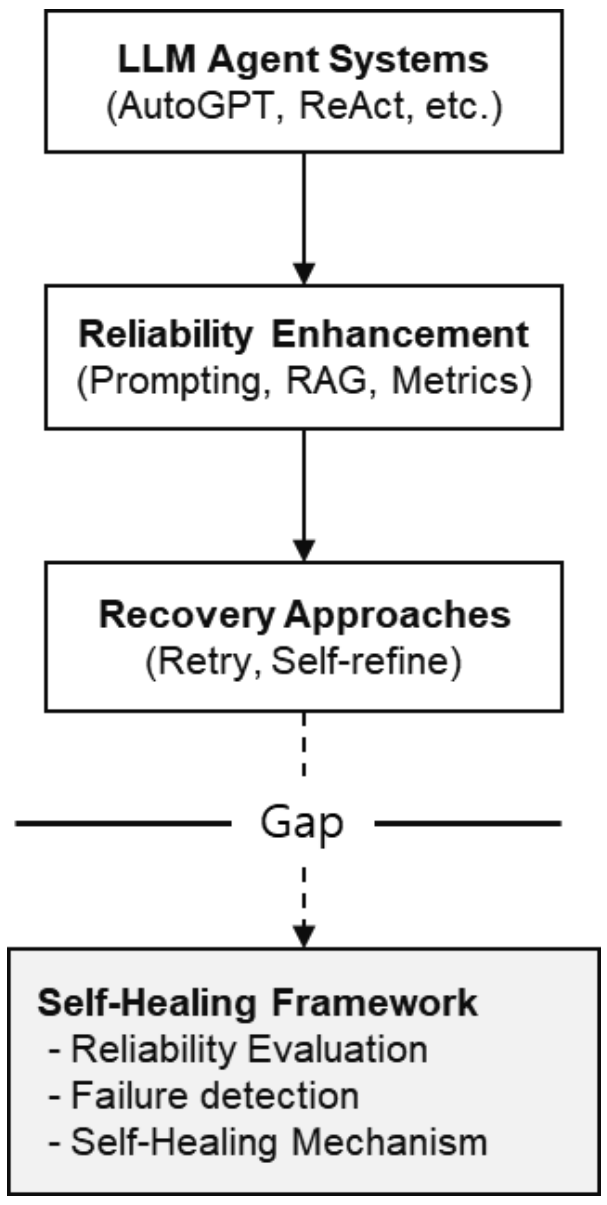

Fig. 1. Comparison of existing approaches and the proposed framework

In contrast, the proposed framework adopts an integrated approach by combining reliability evaluation, failure detection, and self-healing mechanisms into a single system-level architecture. This unified design enables continuous monitoring, dynamic failure detection, and adaptive recovery during runtime, thereby providing a more robust and practical solution for LLM-based autonomous agents.

### 2.1. *LLM-Based Autonomous Agent Systems*

Recent studies have explored LLMs as core components for building autonomous agents capable of reasoning and acting in dynamic environments. ReAct [2] introduces a framework that integrates reasoning and action through interleaved prompting, enabling multi-step decision-making. Similarly, Toolformer [11] demonstrates how language models can learn to use external tools to enhance task performance. HuggingGPT [3] further extends this idea by orchestrating multiple AI models to solve complex tasks collaboratively.

In addition, generative agent frameworks simulate human-like behaviors using LLMs, highlighting the potential of agent-based systems in interactive environments [12]. While these approaches demonstrate strong capabilities, they primarily focus on task performance and lack mechanisms for ensuring reliability and robustness during execution.

### 2.2. *Reliability and Failure Handling in LLM Systems*

Reliability issues in LLMs have attracted increasing attention, particularly in relation to hallucination and reasoning errors [4, 13]. Several approaches aim to improve output correctness through self-consistency decoding, which generates multiple reasoning paths and selects the most consistent answer [6]. Iterative refinement methods, such as Self-Refine [5], further enhance output quality by incorporating feedback loops.

Other studies focus on evaluation and benchmarking of LLM reliability, particularly in code generation and reasoning tasks [14]. These approaches provide useful metrics for assessing model performance but are often limited to offline evaluation settings.

More recent works investigate failure detection in LLM-based agents by analyzing execution patterns and output anomalies [15]. However, these approaches typically focus on detection only and do not provide integrated mechanisms for recovery.

For LLM-based systems to achieve true autonomy and enhance trustworthiness, the need has been raised to move beyond static knowledge bases and evolve toward continuous learning and adaptive agents [16, 17]. In particular, if agents can be equipped with self-evolving capabilities that allow them to independently perform knowledge updates, behavior optimization, and modification of collaborative structures, it becomes possible to build a more robust and highly trustworthy AI ecosystem [18].

### 2.3. *Self-Healing and Adaptive Recovery Mechanisms*

Self-healing systems have been extensively studied in the context of autonomic computing and self-adaptive software systems. The vision of autonomic computing emphasizes self-management capabilities, including self-configuration, self-optimization, and self-healing [7]. Subsequent research has explored architectural frameworks and control mechanisms for adaptive systems [8, 19].

In the domain of LLM-based systems, recent approaches have explored adaptive prompting and retry-based recovery strategies [20]. These methods attempt to correct errors by re-generating outputs or modifying prompts. However, they are often heuristic and lack systematic integration with reliability evaluation.

Overall, existing approaches either focus on improving output quality or handling failures in isolation. There remains a lack of unified frameworks that integrate reliability evaluation, failure detection, and self-healing mechanisms in a coherent manner.

### 2.4. *Summary and Research Gap*

From the above discussion, the following limitations can be identified:

- Lack of integrated frameworks combining reliability evaluation and recovery
- Limited runtime failure detection mechanisms
- Absence of systematic self-healing strategies

To address these gaps, this paper proposes a unified framework that integrates reliability evaluation, failure detection, and self-healing mechanisms for LLM-based autonomous agents.

## 3. Method

This chapter presents the proposed reliability-aware self-healing framework for LLM-based autonomous software agents. The framework is designed to address the limitations identified in the related work by integrating reliability evaluation, failure detection, and self-healing mechanisms into a unified architecture.

### 3.1. *Framework Overview*

The proposed framework aims to enhance the reliability of LLM-based autonomous agents by integrating reliability evaluation, failure detection, and self-healing mechanisms into a unified architecture. Unlike existing approaches that address these aspects independently [5, 15, 20], our framework provides an end-to-end solution for runtime reliability management(figure 2).

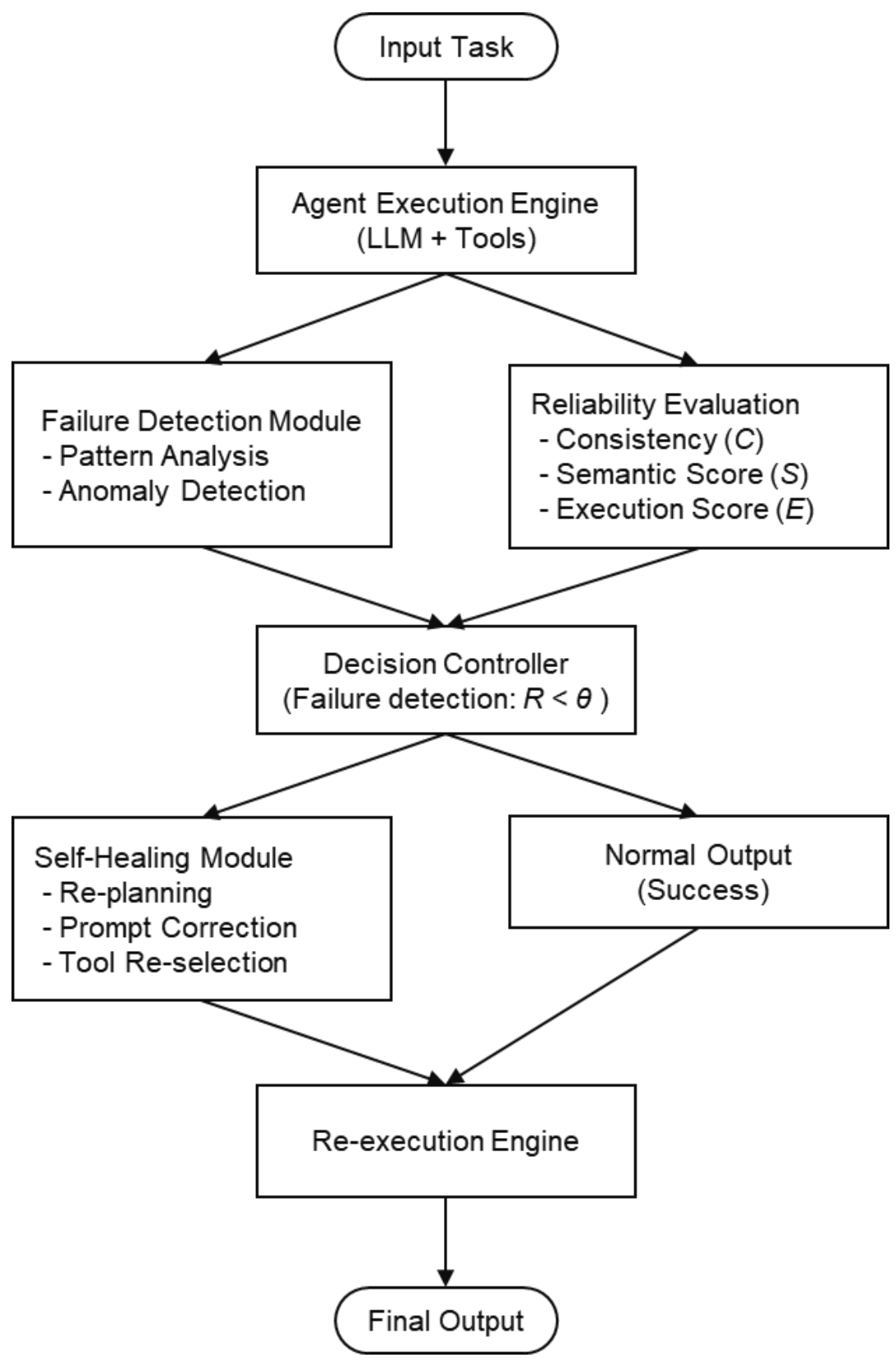

Fig. 2. Architecture of the proposed reliability-aware self-healing framework for LLM-based agents

The proposed framework is composed of several key components that collaboratively enable reliable execution of LLM-based autonomous agents. The process begins with the Task Input, which defines the objective to be executed by the agent. The Agent Execution Engine then performs the task by leveraging LLM capabilities along with external tools or APIs. During execution, the Failure Detection Module continuously monitors the agent’s behavior to identify anomalies or errors. In parallel, the Reliability Evaluation Module quantitatively assesses the quality and consistency of the agent’s outputs.

When a failure is detected or the reliability score falls below a predefined threshold, the Self-Healing Module is activated to apply appropriate recovery strategies, such as re-planning, prompt correction, or tool re-selection. The corrected task is then passed to the Re-execution Engine, which re-runs the process until a satisfactory result is obtained.

Overall, the workflow follows a closed-loop process of Task → Execution → Detection → Evaluation → Healing → Re-execution. This design is inspired by self-adaptive software architectures [8, 19], where monitoring, analysis, and adaptation are tightly integrated to ensure system robustness under dynamic conditions.

Table 1. Comparison of related work and the proposed approach

| **Approach** | **Reliability Evaluation** | **Failure Detection** | **Self-Healing** | **Integration** | **Real-world Applicability** |
|---|---|---|---|---|---|
| ReAct [1] | ✗ | ✗ | ✗ | ✗ | △ |
| Toolformer [11] | ✗ | ✗ | ✗ | ✗ | △ |
| HuggingGPT [3] | ✗ | ✗ | ✗ | ✗ | ○ |
| Self-Refine [5] | △ | ✗ | ○ | ✗ | △ |
| Self-Consistency [6] | △ | ✗ | ✗ | ✗ | △ |
| Autonomic Computing [9] | ○ | ○ | ○ | △ | ○ |
| Existing LLM Recovery | ✗ | △ | △ | ✗ | △ |
| Proposed Method | ○ | ○ | ○ | ○ | ○ |

Note: Table notes.
○: fully supported
△: partially supported
✗: not supported

Table 1 presents a comparison between existing approaches and the proposed framework in terms of key capabilities, including reliability evaluation, failure detection, self-healing, and system integration. As shown in the table, most existing methods support only partial aspects of reliability management. For example, self-refinement approaches improve output quality but lack failure detection and integration, while detection-based methods do not provide recovery mechanisms. In contrast, the proposed framework fully supports all components and provides an integrated solution for end-to-end reliability management. This comprehensive design makes it more suitable for real-world applications where robustness and adaptability are critical.

### 3.2. *Failure Taxonomy*

To systematically handle failures, we define a taxonomy based on common issues observed in LLM-based systems [4, 15]:

- **F1: Hallucination Errors:** Generated outputs that are factually incorrect or unsupported by the context.
- **F2: Execution Errors:** Failures in tool invocation, API interaction, or syntax errors in generated code.
- **F3: Reasoning Inconsistency:** Logical contradictions in multi-step reasoning paths [6]
- **F4: Workflow Propagation Errors:** Cascading failures where an error in one task step causes subsequent dependent tasks to fail.

Unlike prior work that focuses on individual error types [4], our taxonomy provides a unified classification to guide both detection and recovery.

### 3.3. *Reliability Evaluation Model*

We introduce a quantitative reliability model that evaluates agent performance during runtime. The reliability score *R* is defined as:

$$R = \omega_1 C + \omega_2 S + \omega_3 E$$

where:

- *C*: Output consistency (inspired by self-consistency methods [6])
- *S*: Semantic correctness (validated via external knowledge or constraints)
- *E*: Execution success rate
- $\omega_1, \omega_2, \omega_3$ : Weighting coefficients

This model extends existing static evaluation approaches [21] by enabling continuous and dynamic reliability assessment during execution.

Output consistency C measures the similarity among behavioral trajectories $a_1, a_2, \ldots, a_k$ obtained from k independent executions of the same input T. To this end, the Weighted Levenshtein Distance or Jensen-Shannon Divergence can be employed.

$$C = 1 - \frac{2}{T \cdot K(K-1)} \sum_t \sum_{i<j} d_{norm}(a_t^{(i)}, a_t^{(j)})$$

Here, $d_{norm}$ denotes the normalized edit distance, and a value of *C* approaching 1 indicates that the agent's decision-making is stable. Semantic accuracy S is quantified through the text matching rate against a domain-relevant validation database (e.g., RAG). Execution success rate E is defined as the ratio of successful tool and API calls to the total number of attempts.

The parameters $\omega_1, \omega_2, \omega_3$ are weighting coefficients adjusted according to the task domain, and the optimal combination of weights is determined through grid search to maximize overall performance. This model extends existing static evaluation approaches [21] by enabling continuous assessment during execution. Furthermore, this model extends existing evaluation approaches [14] by enabling continuous and dynamic reliability assessment during execution.

where *C* represents output consistency (inspired by self-consistency methods ), *S* denotes semantic correctness (validated via external knowledge or predefined constraints), and *E* is the execution success rate of tools or APIs. The parameters $\omega_1, \omega_2, \omega_3$ are weighting coefficients adjusted based on the task domain. This model extends existing static evaluation approaches [21] by enabling continuous assessment during execution. This model extends existing evaluation approaches [14] by enabling continuous and dynamic reliability assessment during execution.

### 3.4. *Failure Detection Mechanism*

Failure detection is performed using a hybrid approach combining pattern analysis and consistency verification. Failure detection is performed using a hybrid approach combining pattern analysis and consistency verification. First, Execution Pattern Analysis monitors agent behavior to detect anomalies such as repeated tool failures or abnormal execution sequences, motivated by recent work on failure detection in LLM agents [22]. Second, Output Consistency Checking compares multiple reasoning paths to detect contradictions. A failure is triggered when the reliability score $R$ falls below a predefined threshold $\theta$ ($R < \theta$). This enables real-time, threshold-based failure detection.

(1) Execution Pattern Analysis: We monitor agent behavior to detect anomalies such as:

- repeated tool failures
- abnormal execution sequences

This approach is motivated by recent work on failure detection in LLM agents [22].

(2) Output Consistency Checking: We compare multiple outputs or reasoning paths to detect contradictions, following the idea of self-consistency [6].

(3) Threshold-Based Detection: A failure is triggered when $R < \theta$

where $\theta$ is a predefined threshold.

Unlike existing methods that rely on static validation [14], our approach enables real-time failure detection.

### 3.5. *Self-Healing Mechanism*

The self-healing module is designed to automatically recover from detected failures. Autonomic computing, inspired by the self-regulatory capabilities of the human autonomic nervous system, presents a revolutionary paradigm for the design and management of complex computing systems [23, 24].

The Self-Healing module goes beyond simple logical correction by integrating infrastructure-level exceptions into the agent's reasoning loop. For instance, when a specific tool call fails due to a timeout, the system converts the resulting low-level exception message into structured text and delivers it to the agent. The agent then recognizes the situation as an "API response delay" and performs intelligent recovery — such as selecting an alternative tool with lower resource consumption or adjusting the retry interval [25, 26].

Inspired by the principles of autonomic computing [7, 8], we implement three recovery strategies:

(1) Adaptive Re-planning: The agent regenerates a task plan when the current plan fails.
This extends iterative reasoning approaches such as ReAct [2].

(2) Prompt Correction: Prompts are dynamically modified to reduce ambiguity and improve output quality, similar to adaptive prompting methods [20].

(3) Tool Re-selection: Alternative tools or APIs are selected when execution failures occur, extending the tool-usage paradigm introduced in Toolformer [11].

These strategies are adaptively applied according to the detected failure type, enabling context-aware recovery.

For example, upon the occurrence of a hallucination error (F1), the system verifies whether the confidence scores of the three most recent outputs have degraded, and subsequently applies a prompt rewriting strategy. In the case of an execution error (F2), the tool reselection module searches for a suitable alternative API based on evaluation scores. The replanning phase decomposes the task objective into subtasks, excludes any failed subtasks, and performs priority-based re-execution.

### 3.6. *Integrated Algorithm and summary*

The overall process is summarized as follows:

```
Input: Task T
Output: Result R

1: Execute agent on T
2: Compute reliability score R
3: If R < threshold:
4:     Detect failure type
5:     Select recovery strategy
6:     Apply self-healing
7:     Re-execute task
8: Return final result
```

Fig. 3. Integrated Algorithm

Compared to existing approaches, self-refinement methods [5] primarily focus on improving output quality through iterative feedback, but they lack explicit modeling of failure types and do not provide mechanisms for systematic failure handling. Similarly, failure detection approaches [15] are effective in identifying anomalies in LLM-based agent behavior; however, they do not incorporate recovery strategies, limiting their practical applicability in dynamic environments. In addition, adaptive prompting techniques [20] enhance robustness by refining inputs, yet they operate mainly at the prompt level and do not address system-level reliability concerns.

In contrast, the proposed framework provides a unified solution by integrating reliability evaluation, failure detection, and recovery mechanisms into a single architecture. It operates at the system level rather than focusing solely on prompt optimization, enabling comprehensive management of agent behavior. Furthermore, the framework supports runtime adaptation, allowing agents to dynamically detect failures and apply appropriate self-healing strategies during execution, thereby significantly improving overall robustness and reliability.

## 4. Experiment

This section presents the experimental evaluation of the proposed reliability-aware self-healing framework. The objective of the experiments is to assess the effectiveness of the framework in improving the reliability, robustness, and overall performance of LLM-based autonomous agents in realistic task environments.

To this end, we conduct a series of experiments using multi-step workflow scenarios that involve reasoning, tool usage, and iterative task execution. The proposed method is compared with several baseline approaches, including standard LLM agents, retry-based methods, and self-refinement techniques, in order to demonstrate its advantages under different conditions.

The evaluation focuses on key performance metrics such as task success rate, failure detection accuracy, recovery success rate, and execution overhead. In addition, ablation studies and case analyses are performed to examine the contribution of each component and to validate the effectiveness of the integrated design.

### 4.1. *Experimental Setup*

To evaluate the effectiveness of the proposed framework, we conduct experiments in a multi-agent workflow environment that simulates real-world software automation tasks. The experimental setting is designed to reflect practical deployment scenarios of LLM-based agents.

System configuration::

- An LLM-based agent execution engine
- Tool-integrated workflows (e.g., API calls, data processing)
- Monitoring layer for reliability evaluation and failure detection

Task types:

- Multi-step reasoning tasks
- API orchestration workflows
- Document processing pipelines

These tasks are widely adopted in prior studies on LLM-based agent systems [2, 3].

In addition, the experiments employed OpenAI GPT-5 as the base LLM. For each task type (multi-step reasoning, API orchestration, and document processing), 100 test cases were constructed, and a total of 300 task instances were subjected to three repeated experimental runs. Fault injection was implemented by artificially inserting F1–F4 type failures at the execution stage of each task at a random probability of 30%. The statistical significance of the results was verified using the Wilcoxon signed-rank test ($p < 0.05$).

In this experiment, the weighting coefficients were configured according to the task type as follows: for multi-step reasoning tasks, ($\omega_1 = 0.4, \omega_2 = 0.4, \omega_3 = 0.2$); for API orchestration tasks, ($\omega_1 = 0.2, \omega_2 = 0.3, \omega_3 = 0.5$); and for document processing tasks, ($\omega_1 = 0.3, \omega_2 = 0.4, \omega_3 = 0.3$). These configurations reflect the relative importance of output consistency ($C$), semantic accuracy ($S$), and execution success rate ($E$) according to the characteristics of each task type. The threshold value $\theta$ was set to 0.65 based on preliminary experiments.

### 4.2. *Baselines*

We compare the proposed framework with the following baselines:

- **Baseline 1: Standard LLM Agent** - A basic LLM agent with no explicit failure detection or recovery mechanisms, similar to a standard ReAct-style execution.
- **Baseline 2: Retry-Based Approach** - A simple re-execution strategy upon failure detection, commonly used in practical systems.
- **Baseline 3: Self-Refinement Method -** An iterative correction approach using feedback loops, based on methods like Self-Refine.
- Baseline 4: Consistency-Based Method **-** An output selection strategy that leverages self-consistency to choose the most coherent reasoning path.

These baselines represent widely adopted approaches for improving LLM robustness and serve as a comprehensive comparison set for our proposed framework.

### 4.3. *Evaluation Metrics*

We evaluate performance using the following metrics:

- **Task Success Rate (TSR)**: The proportion of successfully completed tasks, indicating the overall effectiveness of the agent.
- **Failure Detection Accuracy (FDA)**: The accuracy of identifying actual failure cases, crucial for timely intervention.
- **Recovery Success Rate (RSR)**: The proportion of detected failures from which the system successfully recovered, demonstrating the efficacy of the self-healing mechanism.
- **Execution Overhead (EO)**: The additional computational time introduced by the framework, assessing its practical feasibility.

These metrics are consistent with prior work on LLM evaluation and reliability analysis.

### 4.4. *Test*

A prototype implementation for validating the proposed framework has been completed, and experiments are currently being conducted across several evaluation dimensions, including TSR, FDA, RSR, EO, and Ablation Study analyses. Preliminary experimental results have shown promising and consistent performance trends. Upon completion of the full validation process, a comprehensive result analysis will be presented to demonstrate the effectiveness and practical applicability of the proposed framework.

## 5. Conclusion & Discussion

This paper presented a novel reliability-aware self-healing framework for LLM-based autonomous software agents, addressing the critical reliability concerns associated with their deployment in complex systems. By integrating a quantitative reliability evaluation model, a hybrid failure detection mechanism, and adaptive self-healing strategies, our framework significantly enhances the robustness and task success rate of LLM agents.

This study contributes to the advancement of reliable LLM-based autonomous systems by providing a comprehensive and integrated solution for runtime reliability management.

Regarding threats to validity, in terms of internal validity, the configuration of weighting coefficients ($\omega_1$, $\omega_2$, $\omega_3$) and the threshold ($\theta$) relies on preliminary experimental results; therefore, recalibration may be required when applied to different task domains. In terms of external validity, since the experiments were conducted in a simulation-based multi-agent environment, additional validation is required for generalization to real-world production environments. In terms of construct validity, as the task success rate (TSR) relies on a binary judgment criterion, more fine-grained performance measurements — such as partial success — warrant consideration in future research.

Furthermore, future work will focus on handling more complex failure types, exploring proactive self-healing mechanisms, and evaluating performance across real-world applications with varying levels of criticality.

## References

[1] Jeong, C., Sim, S., Cho, H., Kim, S., & Shin, B., E2E Process Automation Leveraging Generative AI and IDP-Based Automation Agent: A Case Study on Corporate Expense Processing. *Artificial Intelligence and Applications,* (2025) https://doi.org/10.47852/bonviewAI52026307

[2] Yao, S., Zhao, J., Yu, D., et al, ReAct: Synergizing reasoning and acting in language models. *arXiv preprint arXiv:2210.03629* (2023).

[3] Shen, Y., Song, K., Tan, X., et al, HuggingGPT: Solving AI tasks with ChatGPT and its friends in HuggingFace. *arXiv preprint arXiv:2303.17580* (2023).

[4] Huang, L., Yu, W., Ma, W., Zhong, W., Feng, Z., Wang, H., Chen, O., Peng, W., Feng, H., Qin, B. & Liu, T., A survey on hallucination in large language models: Principles, taxonomy, challenges, and open questions. *ACM Transactions on Information Systems*, 43(2), (2025) 1-55. https://doi.org/10.1145/3703155

[5] Madaan, A., Tandon, N., Gupta, P., Hallinan, S., Gao, L., Wiegreffe, S. & Clark, P., Self-refine: Iterative refinement with self-feedback. *Advances in neural information processing systems (NeurIPS 2023),* 36, (2023) 46534-46594.

[6] Wang, X., Wei, J., Schuurmans, D., Le, Q., Chi, E., Narang, S., Chowdhery, A., & Zhou, D., Self-consistency improves chain of thought reasoning in language models. *arXiv preprint arXiv:2203.11171* (2022).

[7] Kephart, J. O., & Chess, D. M., The vision of autonomic computing. Computer, 36(1), (2003) 41-50. https://doi.org/10.1109/MC.2003.1160055

[8] Salehie, M., & Tahvildari, L., Self-adaptive software: Landscape and research challenges. *ACM transactions on autonomous and adaptive systems (TAAS)*, 4(2), (2009) 1-42. https://doi.org/10.1145/1516533.1516538

[9] Nascimento, N., Alencar, P., & Cowan, D., Self-adaptive large language model (llm)-based multiagent systems. *In 2023 IEEE International Conference on Autonomic Computing and Self-Organizing Systems Companion (ACSOS-C)* (2023) 104-109. https://doi.org/10.1109/ACSOS-58168.2023.00048

[10] Apuri, H., Chinthala, M. M. R., Goel, S., Aurangabadkar, M., & Yepuri, C. Self-Healing Infrastructure: Autonomous LLM Agents for Real-Time Remediation of Configuration Drift and Security Misconfigurations in IaC Deployments. *International Journal of Innovative Technology and Exploring Engineering (IJITEE),* (2026) 25-32. https://doi.org/10.35940/ijitee.D4757.15040326

[11] Schick, T., Dwivedi-Yu, J., Dessì, R., et al. Toolformer: Language models can teach themselves to use tools. *arXiv preprint arXiv:2302.04761* (2023).

[12] Park, J. S., O'Brien, J., Cai, C. J., Morris, M. R., Liang, P., & Bernstein, M. S., Generative agents: Interactive simulacra of human behavior. *In Proceedings of the 36th annual acm symposium on user interface software and technology.* (2023) 1-22. https://doi.org/10.1145/3586183.3606763

[13] Jeong, C., Lee, S., Jeong, S., & Kim, S., A Study on the Framework for Evaluating the Ethics and Trustworthiness of Generative AI. *Artificial Intelligence and Applications*, (2026) https://doi.org/10.47852/bonviewAI62027463

[14] Chen, M., Tworek, J., Jun, H., Yuan, Q., Pinto, H. P. D. O., Kaplan, J., et al,. Evaluating large language models trained on code. *arXiv preprint arXiv:2107.03374* (2021).

[15] Mulian, H., Zeltyn, S., Levy, I., Galanti, L., Yaeli, A., & Shlomov, S., AgentFixer: From Failure Detection to Fix Recommendations in Agentic Systems. *In ACM/IEEE International Conference on Software Engineering* (2026).

[16] Zheng, J., Qiu, S., Shi, C., & Ma, Q., Towards lifelong learning of large language models: A survey. *ACM Computing Surveys*, 57(8), (2025) 1-35. https://doi.org/10.1145/3716629

[17] Yang, Y., Zhou, J., Ding, X., Huai, T., Liu, S., Chen, Q., Xie, Y., & He, L., Recent advances of foundation language models-based continual learning: A survey. ACM Computing Surveys, 57(5), (2025) 1-38. https://doi.org/10.1145/3705725

[18] Jeong, C., A Methodological Framework for Self-Evolving Multi-Agent Systems: Toward Adaptive and Continuous Learning in LLM-Based Architectures. *Research Square*, (2025) https://doi.org/10.21203/rs.3.rs-8139402/v2

[19] Garlan, D., Cheng, S. W., Huang, A. C., Schmerl, B., & Steenkiste, P., Rainbow: Architecture-based self-adaptation with reusable infrastructure. *Computer*, 37(10), (2004) 46-54. https://doi.org/10.1109/MC.2004.175

[20] Ding, D., Xi, W., Ding, Z., & Gao, J., Deep Reinforcement Learning-Driven Adaptive Prompting for Robust Medical LLM Evaluation. *Applied Sciences*, 16(3), (2026) 1514. https://doi.org/10.3390/app16031514

[21] Gupta, A., ReliabilityBench: Evaluating LLM Agent Reliability Under Production-Like Stress Conditions. *arXiv preprint arXiv:2601.06112* (2026).

[22] Ning, K., Chen, J., Zhang, J., Li, W., Wang, Z., Feng, Y.,et al., Defining and Detecting the Defects of Large Language Model-Based Autonomous Agents. *IEEE Transactions on Software Engineering*. (2026) https://doi.org/10.1109/TSE.2026.3658554

[23] Ayomide, A. Y., David, W. O., Oluwanifeni, A. P., Ayomiposi, O. I., Oluwatimilehin, O. M., Oluwapelumi, A. A., et al., Autonomic Computing: Principles, Architecture, Enabling Technologies, Applications, and Future Directions. (2026).

[24] Parashar, M., & Hariri, S., Autonomic computing: An overview. *In International workshop on unconventional programming paradigms.* (2004) 257-269. Berlin, Heidelberg: Springer Berlin Heidelberg.

[25] Miguelañez, C., Designing Self-Healing Systems for LLM Platforms, (2025) https://latitude.so/blog/designing-self-healing-systems-for-llm-platforms

[26] Bantilan, N.. How to Build Self-Healing Agents, (2026) https://www.union.ai/blog-post/how-to-build-self-healing-agents